\documentclass[aps,prl,twocolumn,showpacs,floatfix]{revtex4}
\pdfoutput=1
\usepackage{graphicx}
\usepackage{amssymb}
\usepackage{amsfonts}
\usepackage{amsmath}
\usepackage{dcolumn}

\usepackage{hyperref}

\usepackage{makecell}

\usepackage{color}
\usepackage{stackrel}
\usepackage{latexsym}

\usepackage[normalem]{ulem}
\usepackage{times}
\usepackage{bm}
\usepackage{hyperref}
\usepackage{accents}
\def\bk{{\bf k}}

\begin{document}
\title{Finite temperature Instabilities of 2D Dipolar Bose Gas at Arbitrary Tilt Angle}

\author{Pengtao Shen}
\author{Khandker F. Quader}
\affiliation{Department of Physics, Kent State University, Kent , OH 44242, USA}

\date{\today}

\begin{abstract}

Advances in creating stable dipolar Bose systems, and ingenious box traps have generated tremendous interest.
Theory study of dipolar bosons at finite temperature (T) has been limited. Motivated by these, we study 2D dipolar bosons at arbitrary tilt angle, $\theta$, using finite-T random phase approximation. 
We show that a comprehensive understanding of phases and instabilities at non-zero T can be obtained on concurrently considering dipole strength, density, temperature and $\theta$.
We find the system to be in a homogeneous non-condensed phase that undergoes a collapse transition at large $\theta$, and a finite momentum instability, signaling a striped phase, at large dipolar strength; there are important differences with the T=0 case. At T = 0, BEC appears at critical dipolar strength, and at critical density. Our predictions for polar molecule system, $^{41}K^{87}Rb$, and $^{166}Er$ may provide tests of our results. Our approach may apply broadly to systems with long-range, anisotropic interactions.
	
\end{abstract} 
  

\maketitle

The nature of excitations, phases and instabilities of interacting Bose systems has been a subject of longstanding interest. 
The extraordinary development of the field of ultracold atoms, tremendously advanced by novel experimental techniques,
has, over the past several years, led to intense research. In recent years, there has been considerable interest in systems with long-range and anisotropic
interactions; examples are bosonic and fermionic atoms, and polar molecules experiencing dipolar interactions.

Recent experimental advances in creating stable dipolar bosonic systems, including polar molecules with large electric
dipole moments, have led to vigorous research activities. 
Dipolar Bose-Einstein condensates (BEC) have been realized in chromium~\cite{nature-dg-stable} ($^{52}Cr$), and in lanthanide atoms (such as dysprosium, erbium~\cite{Ferlaino2012}), which have larger magnetic moments. Recent reporting~\cite{ferlaino-roton2018} of observation for the first time of roton mode in dipolar $^{166}Er$ (magnetic moment 7 $\mu_B$) in cigar-shape trap geometry constitute a significant development. Realization of high phase-space density  systems  of polar molecules, such as,  $^{87} Rb^{133}Cs$~\cite{molony2014creation,takekoshi2014ultracold}, $^{41}K^{87}Rb$~\cite{aikawa2009,aikawa2010} hold promise for realization of quantum degeneracy and dipolar BEC. In general, the electric dipole moments of the polar molecules are substantially larger than the magnetic dipole moments of atoms; e.g. 
the RbCs system has sizable electric dipole moment $\sim$ 1.28 Debye. Ingenious box traps constitute another profoundly significant development. There have been box-trap experiments on bosons subjected to contact interaction~\cite{gaunt2013bose,gotlibovych2014observing,lopes2017quasiparticle}; those on dipolar systems are ongoing.

The long-range and anisotropic nature, and a region of attraction, of dipolar interaction can give rise to novel quantum phases, even in dilute systems. A sizeable body of theory work, based on
Monte Carlo and Bogoliubov-de Gennes (BdG) methods, exist at at zero temperature (T=0).  The existence of roton mode and density-wave phase has been found in BdG calculations studying the property of BEC ground state~\cite{Fischer,santos2003,fedorov2014two,lu2015prl,SQJP18}. Solid and stripe like crystal phases have been  predicted in Monte Carlo simulation~\cite{macia2014phase,Bombin2017}. Such stripe phase of the dipolar Bose system provides a promising candidate~\cite{Boninsegni2012rmp,Wenzel2017pra,Tanzi2019prl,Bottcher2019prx,Chomaz2019prx} for an intrinsic supersolid without the presence of defects as described by the Andreev-Lifshitz mechanism~\cite{cinti2014defect}. However, theoretical study of 2D dipolar boson gas at finite temperature has been limited.

A purely dipolar 3D system is usually unstable towards collapse due to the attractive component of the interaction.  A trap helps to stabilize the system; this depends strongly on trapping geometry~\cite{nature-dg-stable, eberlein2005exact}. In 2D, the stability issue of dipolar bosons may be richer~\cite{sun2010spontaneous,yamaguchi2010density}.
There have been studies~\cite{mueller2000, van2002dilute} of density response in 2D and 3D Bose gas with attractive constant interaction, using random phase approximation (RPA). Similar study of dipolar bosons in a cylindrical trap at finite temperature shows that a pancake geometry trap stabilizes the system~\cite{bisset2011}. 
That brings up the question whether purely 2D dipolar bosons are stable at finite temperature. 

In this paper, we present results of our study of a 2D dipolar Bose system at {\it non-zero temperatures}, using finite-temperature RPA~\cite{mueller2000, bisset2011}.
A key point of the paper is that a broad perspective on the phases and instabilities of the system at {\it finite temperature} can be attained by considering several tunable knobs: density, temperature, interaction strength and the orientation of dipole moments to an external field, i.e. tilt angle. We construct several informative phase diagrams based on our
study of dipolar length (strength) versus dipole tilt angle at a given temperature; RPA critical temperature and critical density versus dipolar length for a given tilt angle, and critical temperature versus critical density for a given tilt angle. In particular, at finite temperature, we find the system to be in a homogeneous non-condensed phase that undergoes a collapse transition at large tilt angles, and a finite momentum instability, signaling a striped phase, at sufficiently large values of dipolar coupling strength; there are substantive differences with the T = 0 case.  
The linear $q$ dependence of 2D dipolar interaction is manifested in a new density wave instability in a broad regime similar to 2D dipolar fermions~\cite{sun2010spontaneous, yamaguchi2010density}. 
As $T \rightarrow 0$, for sufficiently small tilt angles, BEC appears at critical dipolar strength, and at critical density, values of which depend on other system parameters.

While our results should apply generally to 2D dipolar bosons at arbitrary temperature, the specific predictions based on parameters of the polar molecule system, $^{87} Rb^{133}Cs$, can serve as a test of 
our results. We discuss the effects of a harmonic trap and an additional contact interaction; however, key physics is captured in our consideration of 
a homogeneous 2D Bose system. Also, aforementioned box traps makes study of homogeneous systems directly relevant and testable. 

\begin{figure}[h]
	\includegraphics[width=16pc]{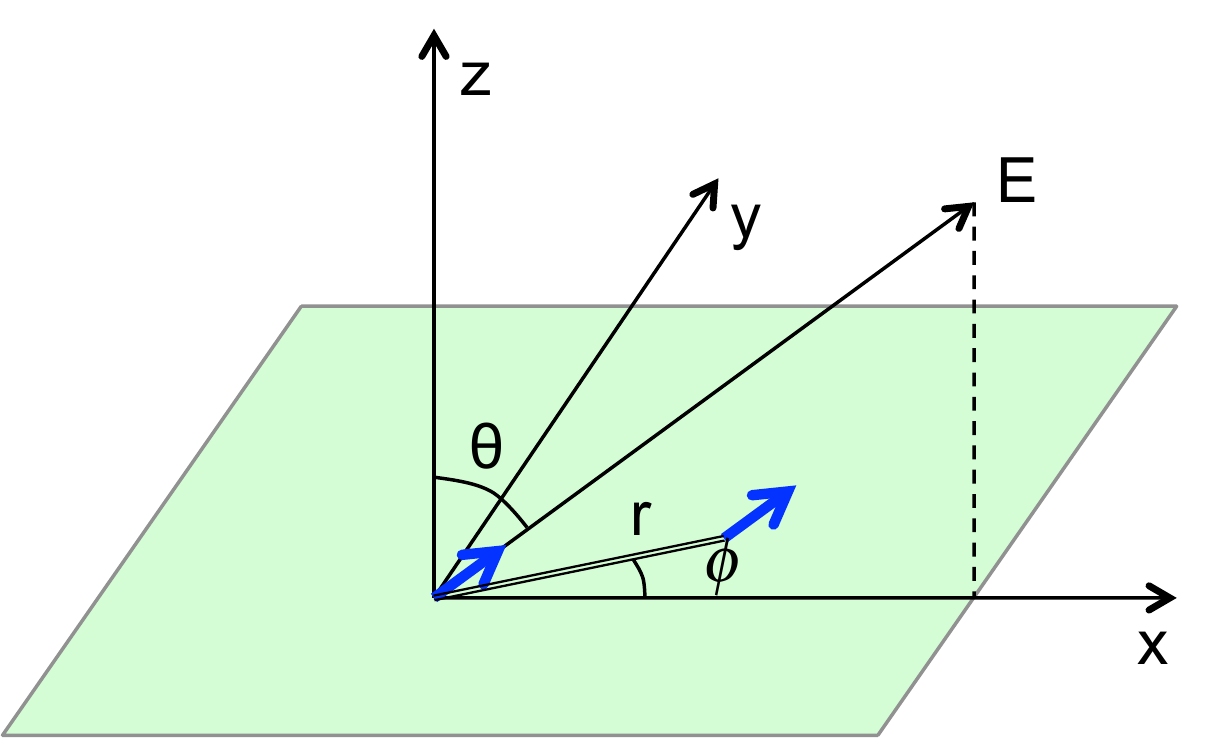}\hspace{4pc}%
	\caption{2D Dipoles in x-y plane with tilt angle $\theta$ that defines the direction of electric/magnetic field E relative to the z-direction. $\phi$ is the angle in x-y plane, relative to x-direction.}
	\label{geometry}
\end{figure}

We consider a gas of dipolar bosons of mass $m$ and electric or magnetic dipole moment $d$. The dipoles are confined by in x-y plane and the dipole moments are aligned by an external electric field $\textbf{E}$ 
or magnetic field $\textbf{B}$, subtending an angle $\theta$ with respect to the z axis as shown
in Fig. \ref{geometry}.

To explore the stability of the system, we consider the static density-density correlation function $\chi(q)\equiv\chi(q,w=0)$, which determines the stability of a system against density fluctuations.
When $\chi(q)$ becomes positive at wave vector q, the system has density wave instability and may undergo a transition to the striped phase~\cite{sun2010spontaneous,yamaguchi2010density}; if it becomes positive as $q\to0$, the system will develop negative compressibility and collapse. 

In standard RPA, the density-density correlation function can be diagrammatically expanded in terms of the bare response function  $\chi_0$ (see Supplemental Materials, Fig. S1~\cite{supp}).
For the stability condition against density fluctuations of finite momentum, the direct scattering of particle-hole excitations dominates over exchange scattering because of the linear momentum dependence of 
$V_{2D} (q)$~\cite{bisset2011,yamaguchi2010density}. So, we neglect the exchange scattering of particle-hole excitations. Also, since in 2D, there is no BEC at finite temperature, 
there is no contribution from the condensate in the RPA response. Then,

\begin{eqnarray}\label{response}
\chi(q,\omega)=\frac{\chi_0(q,\omega)}{1-V(q)\chi_0(q,\omega)}
\end{eqnarray}
with
\begin{eqnarray}\label{barebubble}
\chi_0(q,\omega)=\int \frac{d \bk}{(2\pi)^d} \frac{f(k-q/2)-f(k+q/2)}{\hbar\omega-(\varepsilon_{k+q/2}-\varepsilon_{k-q/2})}
\end{eqnarray}
where $\varepsilon_k=\hbar^2 k^2/{2m}$ is the free particle kinetic energy, and $f(q)$ is the Bose distribution function, with chemical potential $\mu$. For non-interacting bosons, $\chi=\chi_0$ and 
always negative, so the system is stable. For interaction with an attractive channel, the system can become unstable depending on the interaction.

To proceed, we first need to evaluate the finite temperature bare response function, Eq. (\ref{barebubble}). The asymptotic behavior is given by (for details, see Supplemental Materials, Bare Bubble Calculation and Fig. S2~\cite{supp}:\\
In the $q\lambda_T\ll1$ region, 
\begin{eqnarray}\label{chi0s}
\chi_0(q,T)=-\frac{m}{2\pi\hbar^2}\frac{1}{e^{-\beta\mu}-1}(1+O(q\lambda_T)^2)
\end{eqnarray}
where $\lambda_T=\sqrt{\frac {2\pi\hbar^2\beta}{m}}$ is the thermal de Broglie wavelength and $\beta = 1/k_BT$.

In the $q\lambda_T\gg1$ region, the behavior is temperature-independent,
\begin{eqnarray}\label{chi0l}
\chi_0(q)=-\frac{4nm}{\hbar^2q^2}
\end{eqnarray}

We calculate RPA responses, taking bare response function to be that of a noninteracting system, and using the non-interacting gas to calculate the chemical potential $\mu(T,n)$.  
For an ideal two dimensional boson gas, the 2D density is 
\begin{eqnarray}
n&=&\int \frac {d\bk}{(2\pi)^2} \frac{1}{e^{\beta(\varepsilon_k-\mu)}-1}\nonumber\\
&=&\lambda^{-2}_T g_1(e^{\beta\mu})
\end{eqnarray}
$g_v(z)=\sum_j z^j/j^v$ is the polylogarithm function. 
The chemical potential $\mu$ is
\begin{eqnarray}
\mu=\frac1\beta \ln[1-\exp(-n\lambda_T^2)]
\end{eqnarray}
As temperature decreases, $\chi_0(q)$ increases. In the classical limit, $\beta\mu\gg-1$, Eq. (\ref{chi0s}) become $\chi_0(0)=-n/T$, independent of Bose statistics. In the quantum limit, $\beta\mu\ll-1$, Eq. (\ref{chi0s}) become $\chi_0(0)=-\frac{m}{2\pi\hbar^2}e^{n\lambda^2_T}$.
Since there is no upper limit for $g_1(z)$ at zero chemical potential, there is no BEC in 2D at finite temperature. In the limit of zero temperature,
$\lim\limits_{T\to 0}\mu=0$, and $\lim\limits_{T\to 0}n(k)=n\,\delta(k)$. The limiting behavior at zero temperature is the BEC state; Eq. (\ref{chi0s}) become $\chi_0(0,0)=-\infty$ and Eq. (\ref{chi0l}) becomes the response function of ideal bosons for all momenta at T= 0. 

It is convenient to look at the inverse of static density-density correlation function $\chi(\textbf{q})$, given by:
\begin{eqnarray}
\frac1{\chi(\textbf{q})}=\frac1{\chi_0(\textbf{q})}-V(\textbf{q})
\end{eqnarray}
where $V(\textbf{q})$ is the dipole-dipole interaction (DDI), given by $V(\textbf{q})=V_s+V_l(\textbf{q})$, with
\begin{eqnarray}
\begin{aligned}
&V_s=2\pi d^2 \frac{P_2(\cos\theta)}{r_c}\\
&V_l(\textbf{q})=-2\pi d^2q({\cos^2}\theta-{\sin^2}\theta {\cos^2}\phi)
\end{aligned}
\end{eqnarray}
where $r_c$ is a short range cut off. In quasi-2D geometry it depends on the trapping size in z direction~\cite{Fischer}. The first term $V_s$ is momentum independent and acts like a short range interaction. The second term depends linearly on the magnitude of momentum.  In the y-direction ($\phi=\pi/2$),  the interaction is the most attractive; therefore, the instability happens at momentum in y direction. 
Three distinct cases may be noted: 

1. $\frac1{\chi_0(q)}-V(q)<0$ everywhere; the system is in uniform normal stable phase.

2. $\frac1{\chi_0(q)}-V(q)>0$ at finite q; system undergoes density wave instability and has striped phase.

3. $\frac1{\chi_0(q)}-V(q)>0$ at q=0; the system has negative compressibility and will collapse.

\begin{figure}[h]
	\begin{minipage}[b]{1\linewidth}
		\centering		\includegraphics[width=\linewidth]{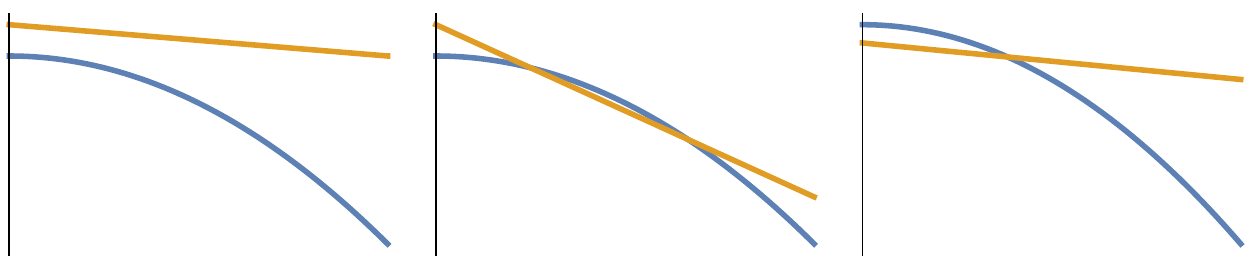}
	\end{minipage}
	\caption{(color online) Schematic illustration for three cases of response function discussed in text, curved line is $1/\chi_0(q)$, straight line is V(q)}
	\label{chivsV}
\end{figure}

Fig. \ref{chivsV} shows schematically the behavior of $\frac1{\chi_0(q)}$ and $V(q)$ for the three cases.  $\frac1{\chi_0(q)}$ is approximately quadratic in q with intercept $\frac1{\chi_0(0)}\leq0$   ; $V(q)$ is linear in q with negative slope and positive/negative intercept depending on the tilt angle. Note that instability for dipolar system at finite q is possible in 2D, but not in 3D~\cite{bisset2011}, because $V(q)$ is independent of the magnitude of momentum in 3D~\cite{sun2010spontaneous}.

The behavior of the response function for the 2D dipolar Bose  gas at non-zero T can be generally categorized into two regimes with respect to tilt angle $\theta$:

A) $\theta<\cos^{-1}{\frac1{\sqrt3}}$: the short range interaction $V(q=0)$ is positive and $\frac1{\chi_0(0)}\le0$, $\frac1{\chi_0(0)}-V(0)\leq0$ is always satisfied, thus it is impossible for the system to collapse. However, the instability condition could be satisfied at finite $q_y$ with a sufficiently strong dipole interaction strength, describable by dipolar length, $a_{dd}=md^2/\hbar^2$.  Thus the system is unstable against a density fluctuation with wave-vector $q_y$. That indicates a transition from the normal phase to a striped phase. 

B) $\theta>\cos^{-1}{\frac1{\sqrt3}}$: in this region, the short range interaction becomes negative.
At zero temperature, the bare response function diverges at q=0 and $\frac1{\chi_0(0)}=0$, so the system cannot support any attractive short range interaction. Thus it always collapses at T= 0,
as in the BdG approach. However, {\it at finite temperature}, the bare response function has a non-zero negative value at q=0. As a result, it can support attractive short range interaction that is sufficiently small. The system first undergoes transition from normal phase to a striped phase; then collapses as $a_{dd}$ increases further. At large tilt angles close $\pi/2$, the long-range interaction becomes zero, and the total interaction is dominated by the negative short range part of the dipolar interaction. Then the system goes from the normal to the collapsed phase without going through an intermediate striped phase. 

\begin{figure}[h] 
	\centering
	\begin{minipage}[b]{0.48\linewidth}
		\centering
		\includegraphics[width=\linewidth]{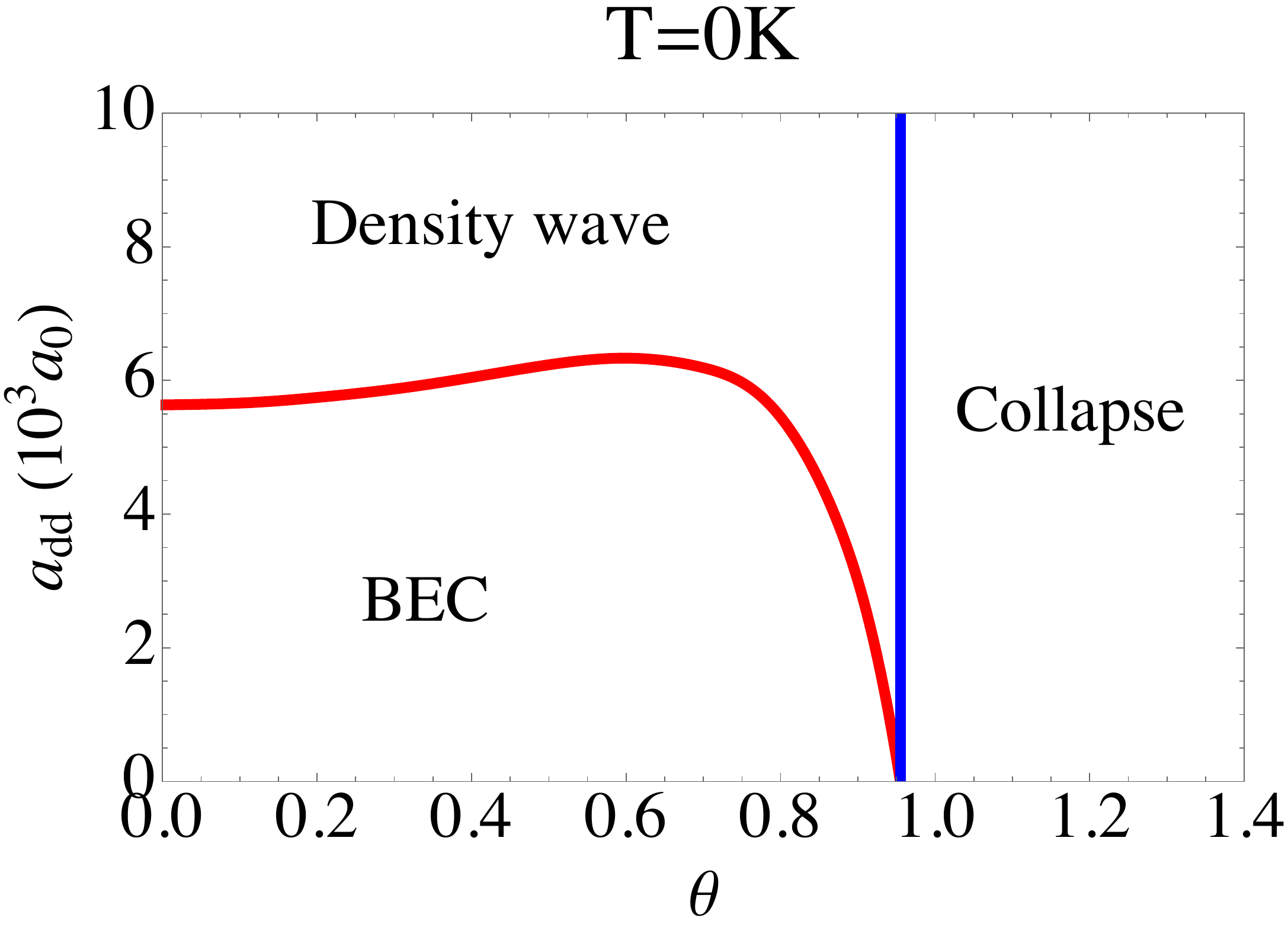}
	\end{minipage}
	\hfill
	\begin{minipage}[b]{0.48\linewidth}
		\centering
		\includegraphics[width=\linewidth]{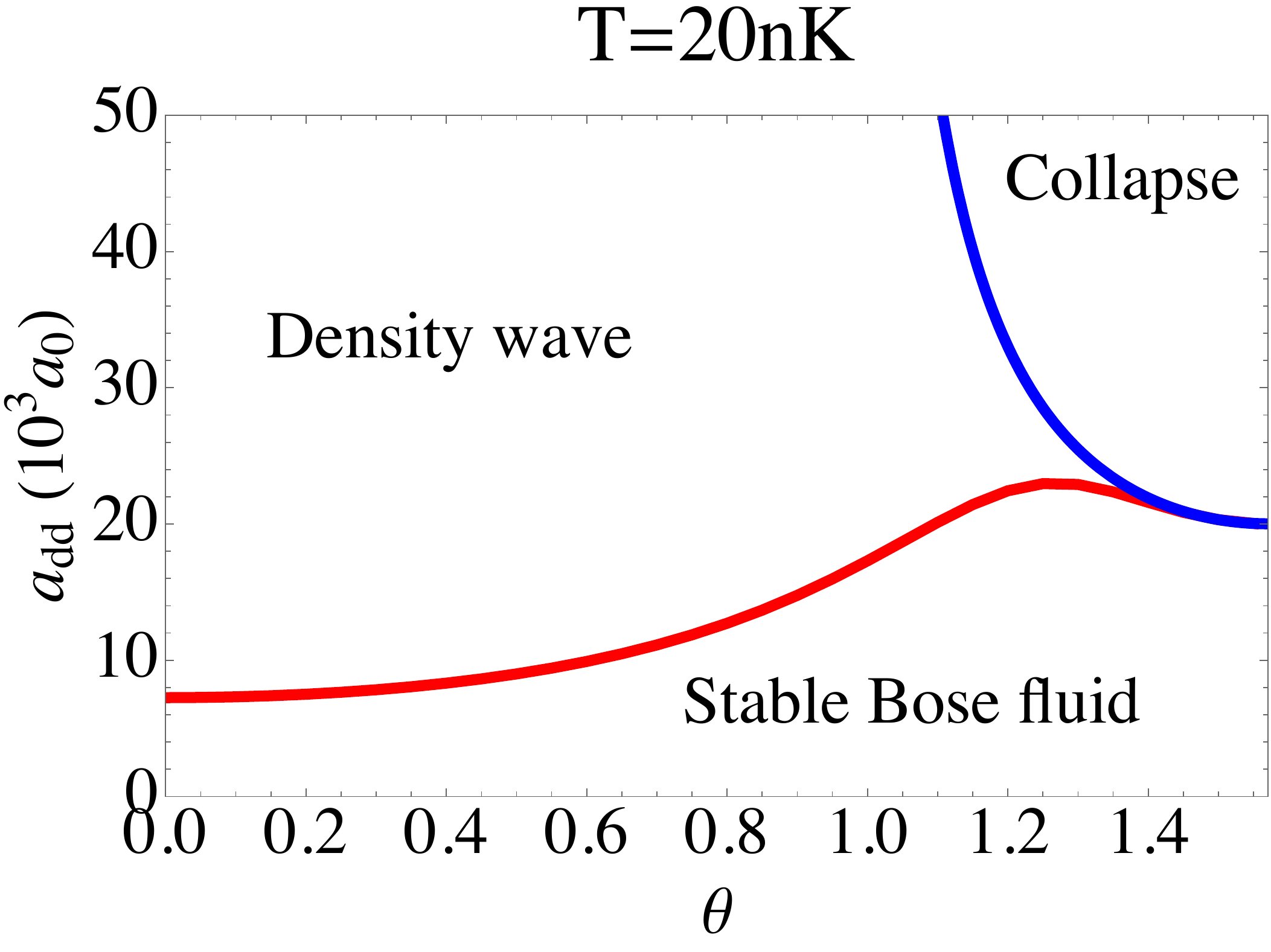}
	\end{minipage}
	\caption{(color online) Dipole interaction strength $a_{dd}$ versus tilt angle $\theta$ for a system of polar molecule $^{87}Rb^{133}Cs$ at T=0 and T=20K.  $a_{dd}$ is in units of the Bohr radius $a_0$. Red and blue lines are lines of density wave and collapse instabilities respectively.}
	\label{add-theta}
\end{figure}

For fixed density and temperature, the dipole interaction strength $a_{dd}$ can be changed via the strength of the external field,  and dipole's tilt angle $\theta$ by varying the direction of external field. In Fig. \ref{add-theta}, we show the calculated phase diagram for critical dipole interaction strength $a_{dd}$ versus the tilt angle $\theta$ for a system of polar molecule such as $^{87}Rb^{133}Cs$ at T=0 and T=20 nK. We choose density to be  $n=10^{12}m^{-2}$ and take the cut-off to be $r_c=10^4a_0$; mass of $^{87}Rb^{133}Cs$ is $m=220u$ (unified atomic mass unit). At T= 0 K, the system goes from stable BEC to density wave instability as dipole strength $a_{dd}$ increases when $\theta<0.955$; the system collapses for any dipole strength when $\theta>0.955$. On the other hand, at T= 20 nK, density wave instability appears as $a_{dd}$ increases, even for tilt angle $\theta>0.955$. The system eventually collapses as $\theta$ is increased further. That the collapse occurs at a larger value of $\theta_c$  compared to the T=0 case may be understood on noting the interplay between the DDI and the bare T-dependent response, $\frac1{\chi_0(T)}$ in the RPA response function. 

\begin{figure}[ht]
	\centering
	\begin{minipage}[b]{0.48\linewidth}
		\includegraphics[width=\linewidth]{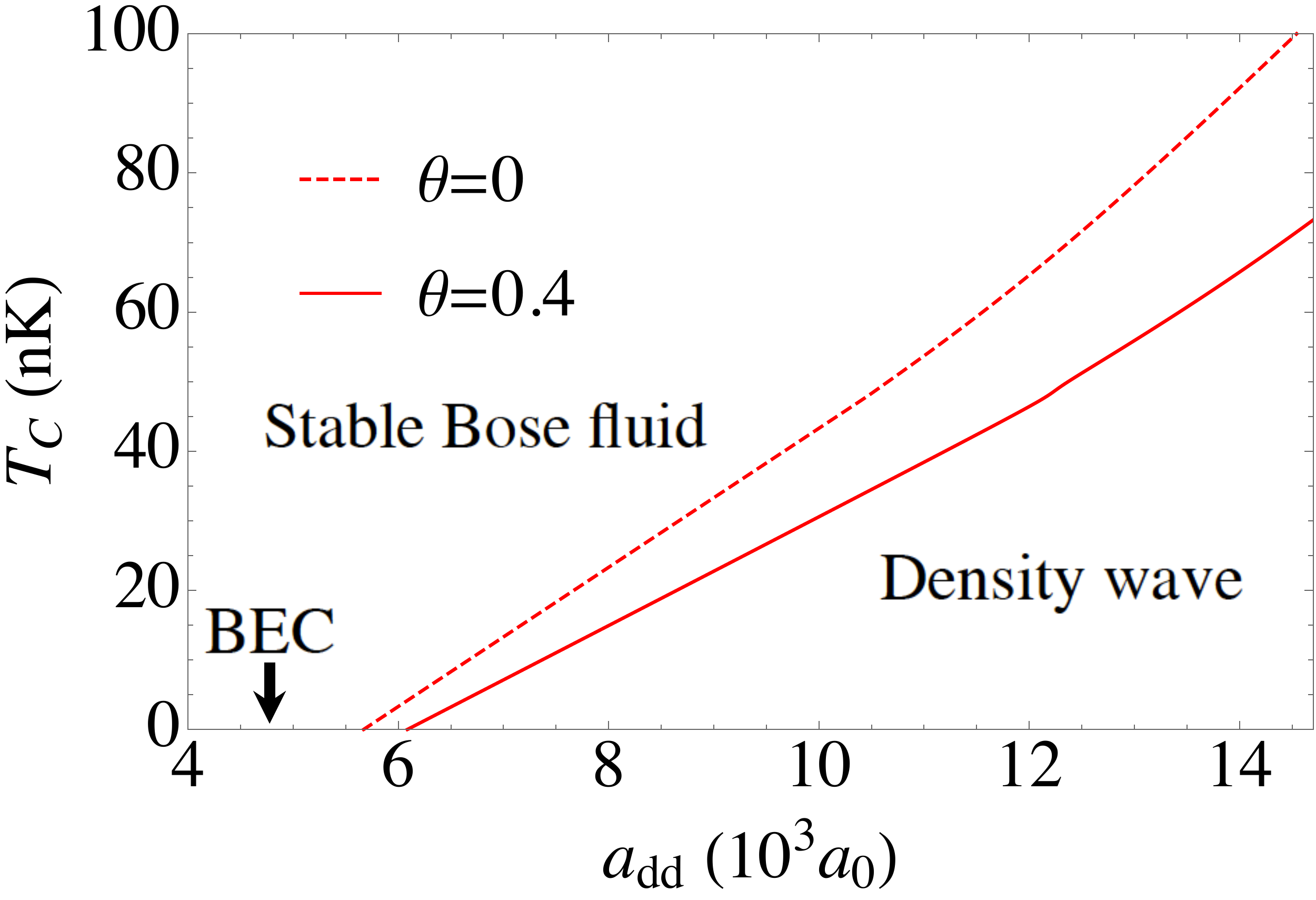}
	\end{minipage}
	\hfill
	\begin{minipage}[b]{0.48\linewidth}
		\centering
		\includegraphics[width=\linewidth]{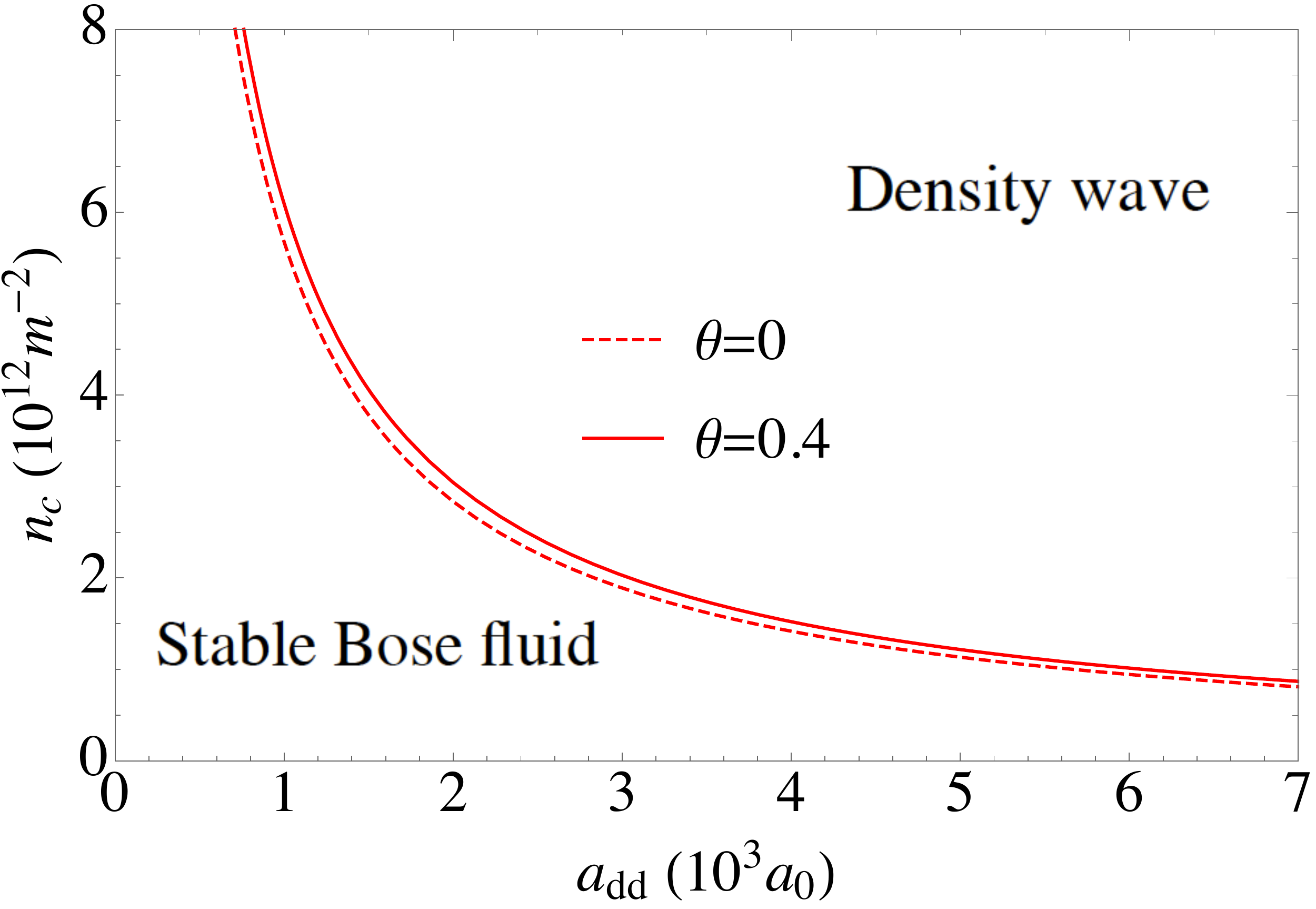}
	\end{minipage}
	\caption{(color online) Critical temperature and critical density for density wave instability of $^{87}Rb^{133}Cs$ at tilt angles $\theta=$ 0 and 0.4. Left: $T_c$ vs $a_{dd}$, for $n=10^{12}m^{-2}$. Right: $n_c$ vs $a_{dd}$ at T=10 nK.}
	\label{TcNc-add}
\end{figure}

Study of critical temperature $T_c$ and critical density $n_c$, albeit within RPA, provide another perspective on phases and density wave instability in the system. We first 
calculate $T_c$ and critical density $n_c$, each as a function of $a_{dd}$, for fixed tilt angle, $\theta$. Fig. \ref{TcNc-add}, shows our results for the physical system, 
$^{87}Rb^{133}Cs$ for $\theta=$ 0 and 0.4. $T_c$  and $n_c$ are calculated using the condition $\frac1{\chi(q,T)}=0$. For calculation of  $T_c$, we choose the system density to be $10^{12}m^{-2}$, and for $n_c$, 
the system temperature to be 10 nK. In the context of density wave instability, a key point here is that the critical temperature and critical density behave opposite to each other with increasing dipole strength, i.e. $T_c$ increases, and $n_c$ decreases when dipole strength is increased (see Fig. \ref{TcNc-add}). The  terminating point ( $T = 0$) of the $T_c$ curves 
signifies the onset $a_{dd}$ for BEC; this is dependent on the tilt angle. 
 
\begin{figure}[t]
	\centering
	\begin{minipage}[b]{0.48\linewidth}
		\centering
		\includegraphics[width=\linewidth]{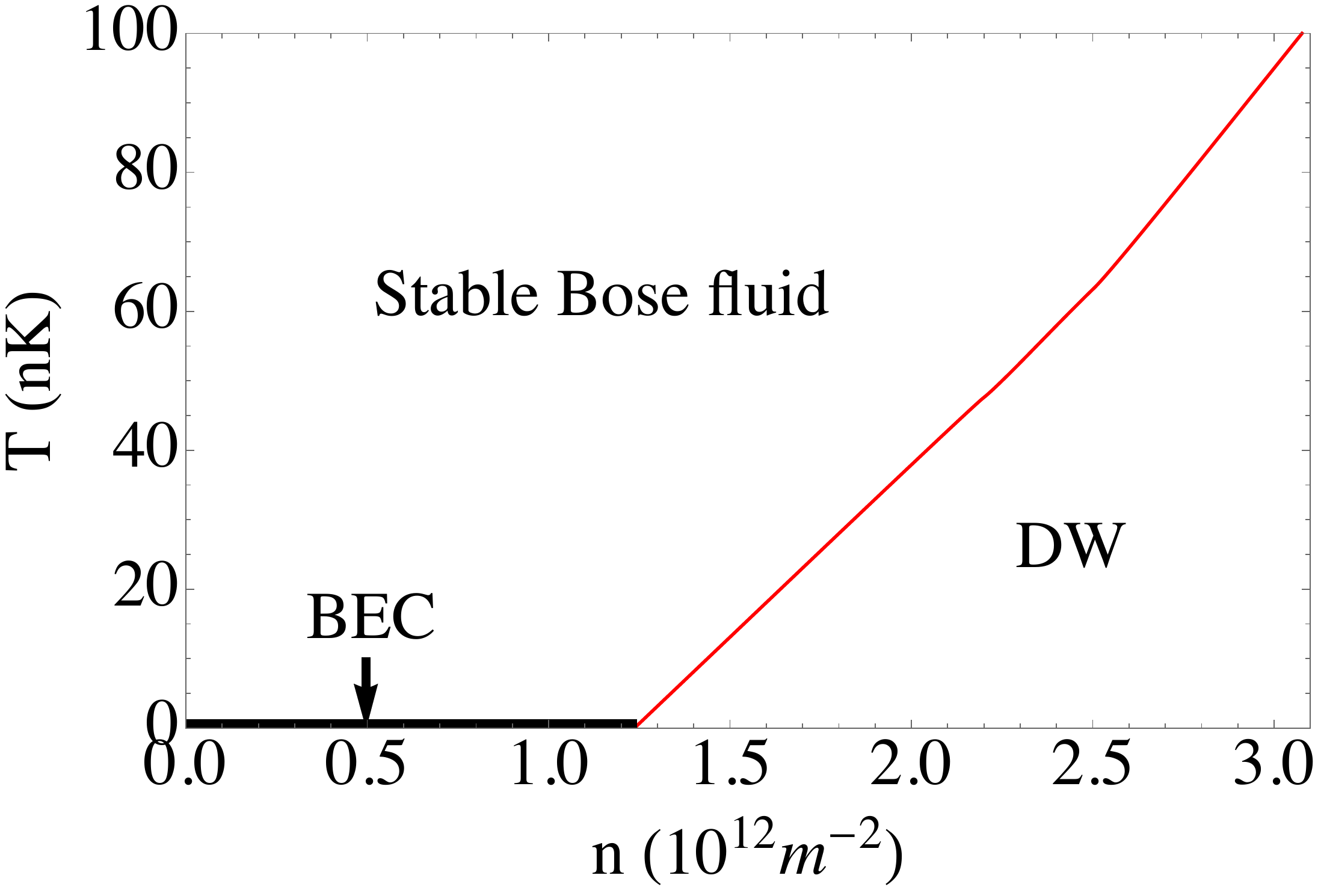}
	\end{minipage}
	\hfill
	\begin{minipage}[b]{0.48\linewidth}
		\centering
		\includegraphics[width=\linewidth]{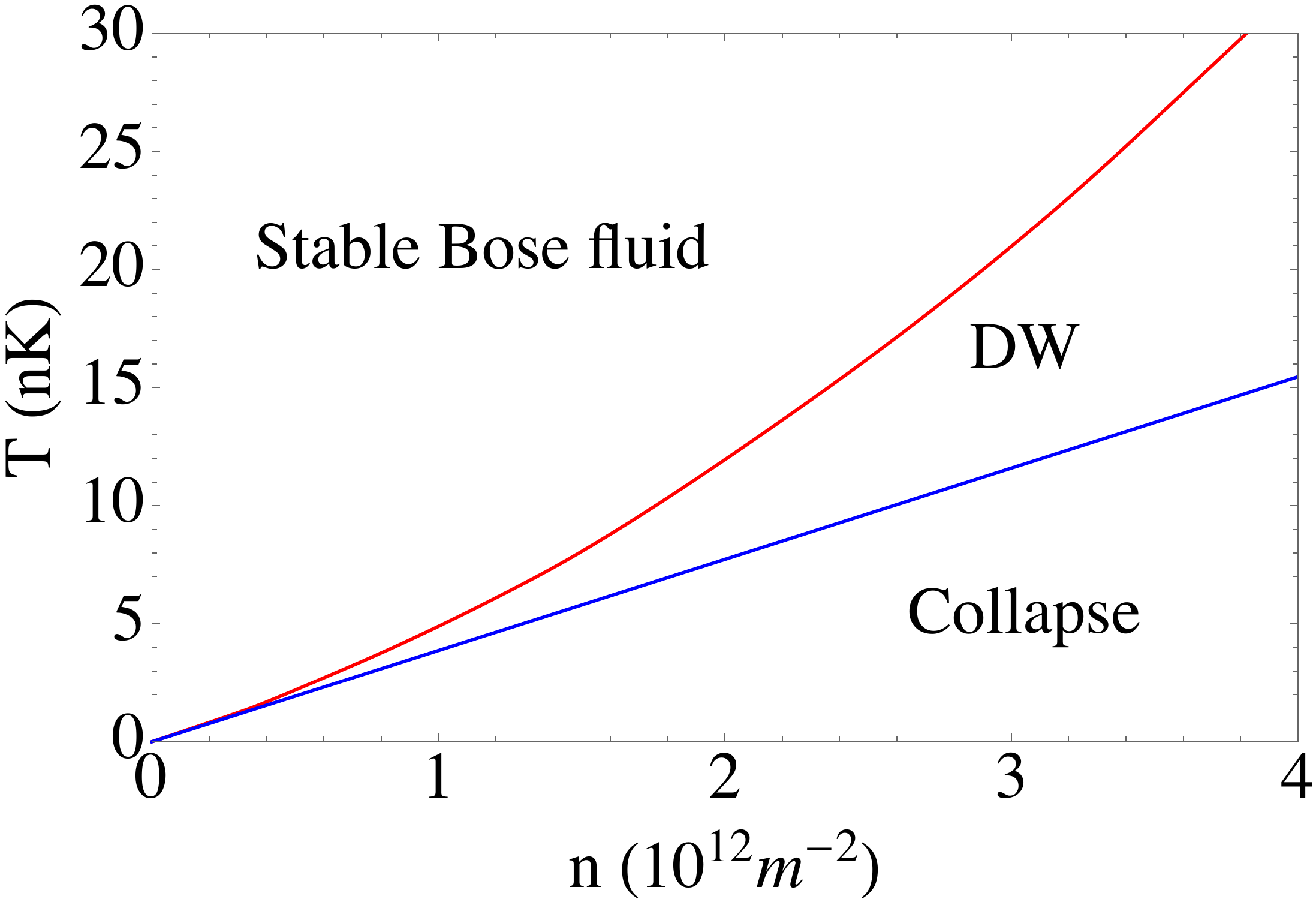}
	\end{minipage}
	\begin{minipage}[b]{0.48\linewidth}
		\centering
		\includegraphics[width=\linewidth]{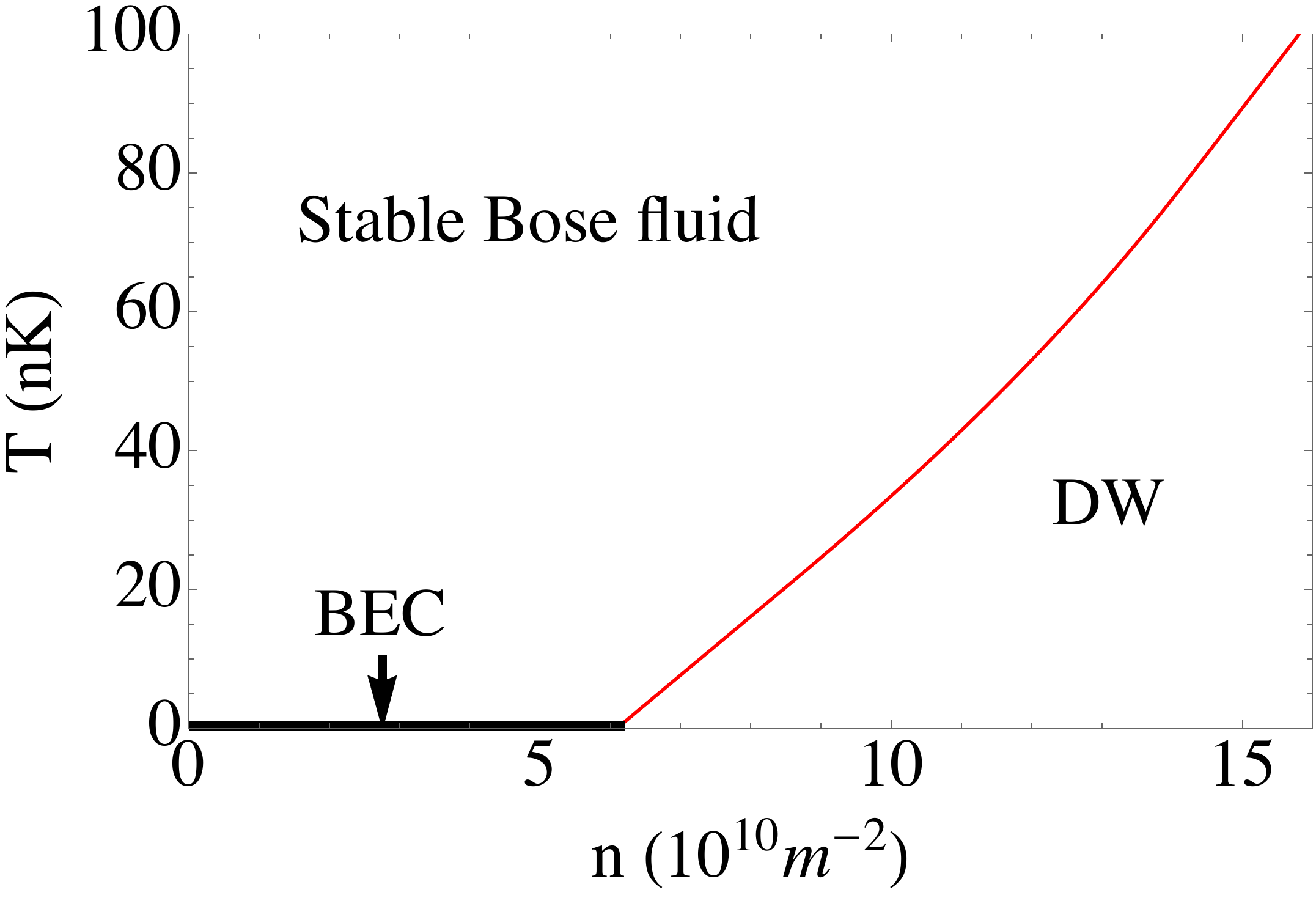}
	\end{minipage}
	\hfill
	\begin{minipage}[b]{0.48\linewidth}
		\centering
		\includegraphics[width=\linewidth]{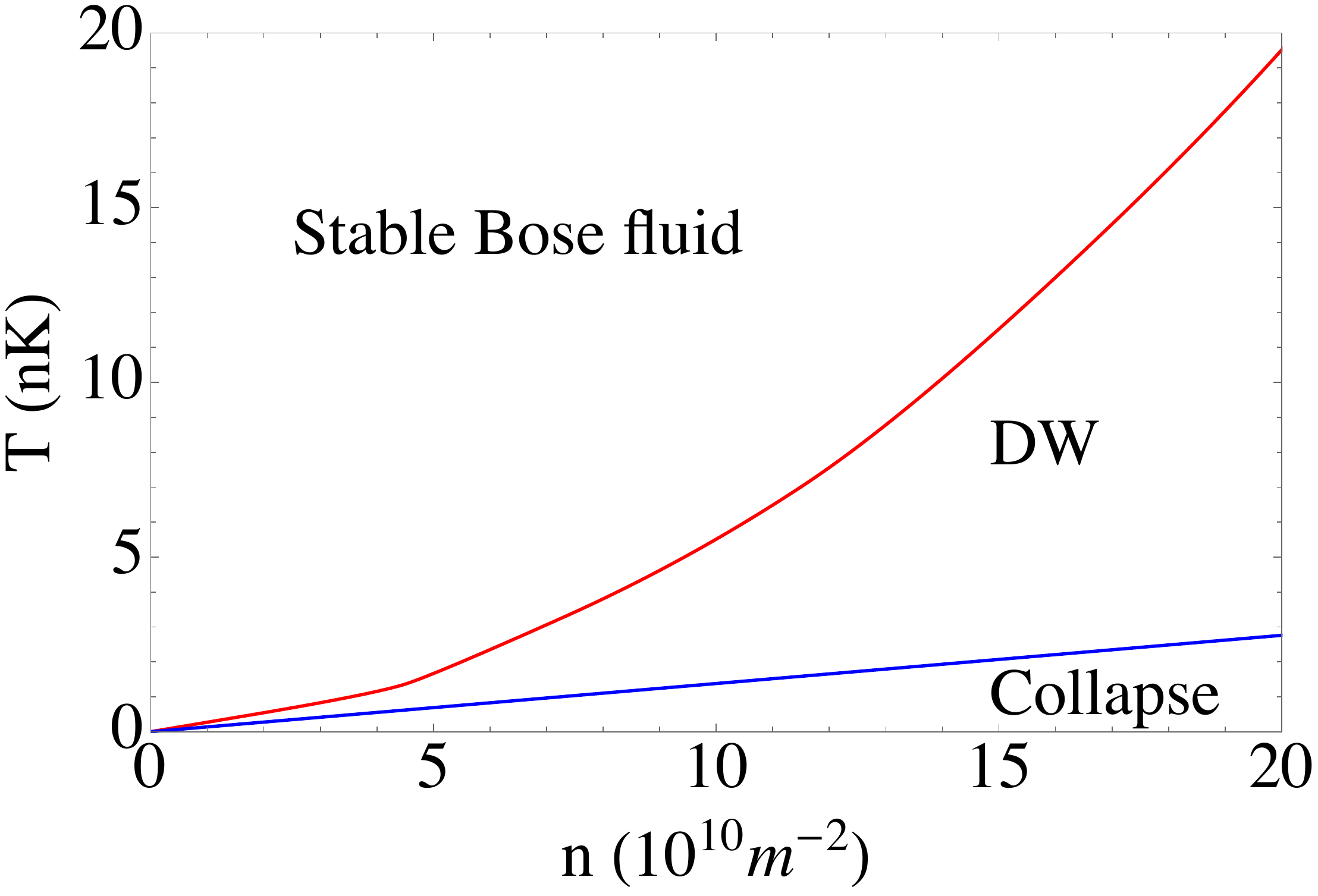}
	\end{minipage}
	\begin{minipage}[b]{0.48\linewidth}
		\centering
		\includegraphics[width=\linewidth]{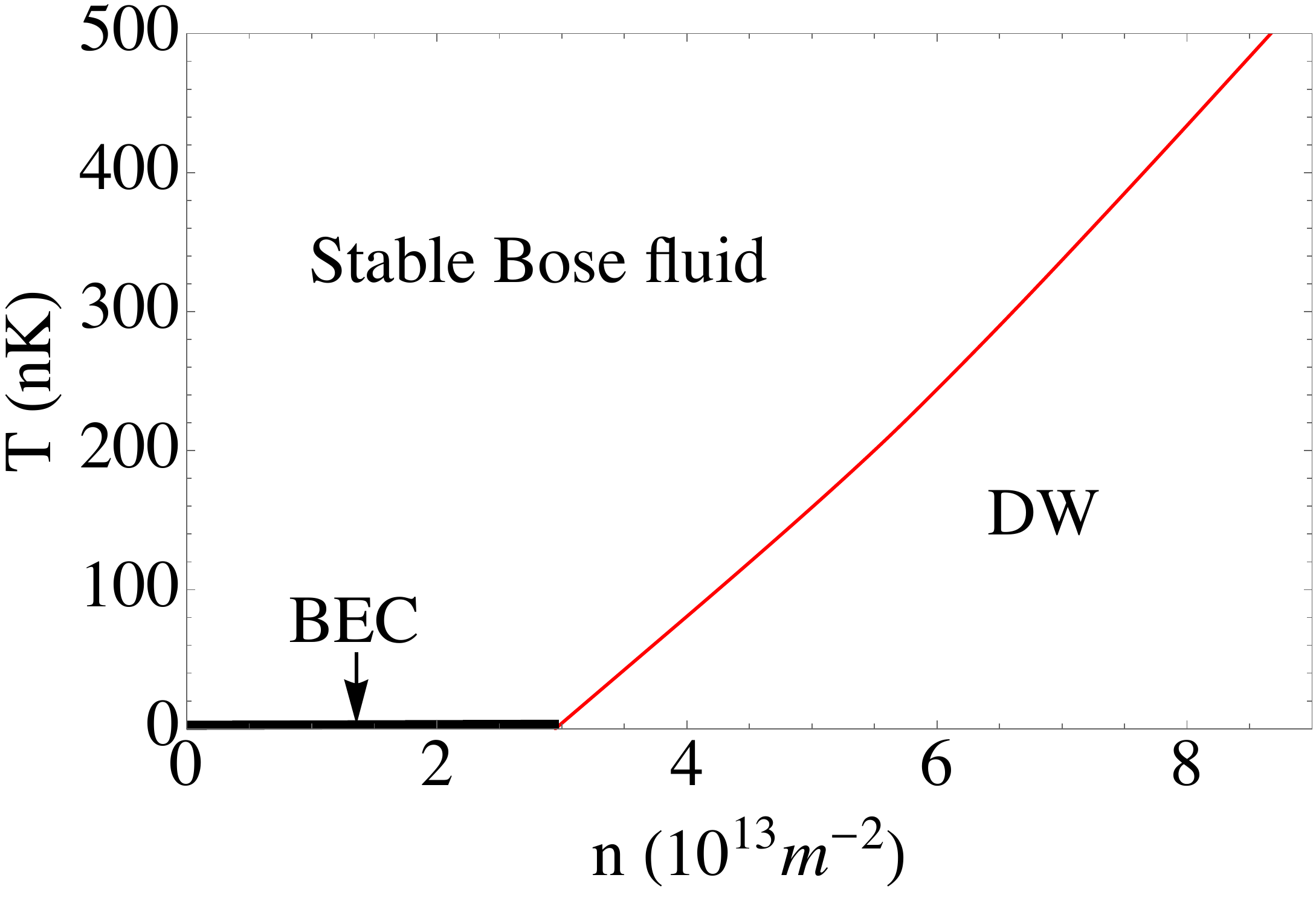}
	\end{minipage}
	\hfill
	\begin{minipage}[b]{0.48\linewidth}
		\centering
		\includegraphics[width=\linewidth]{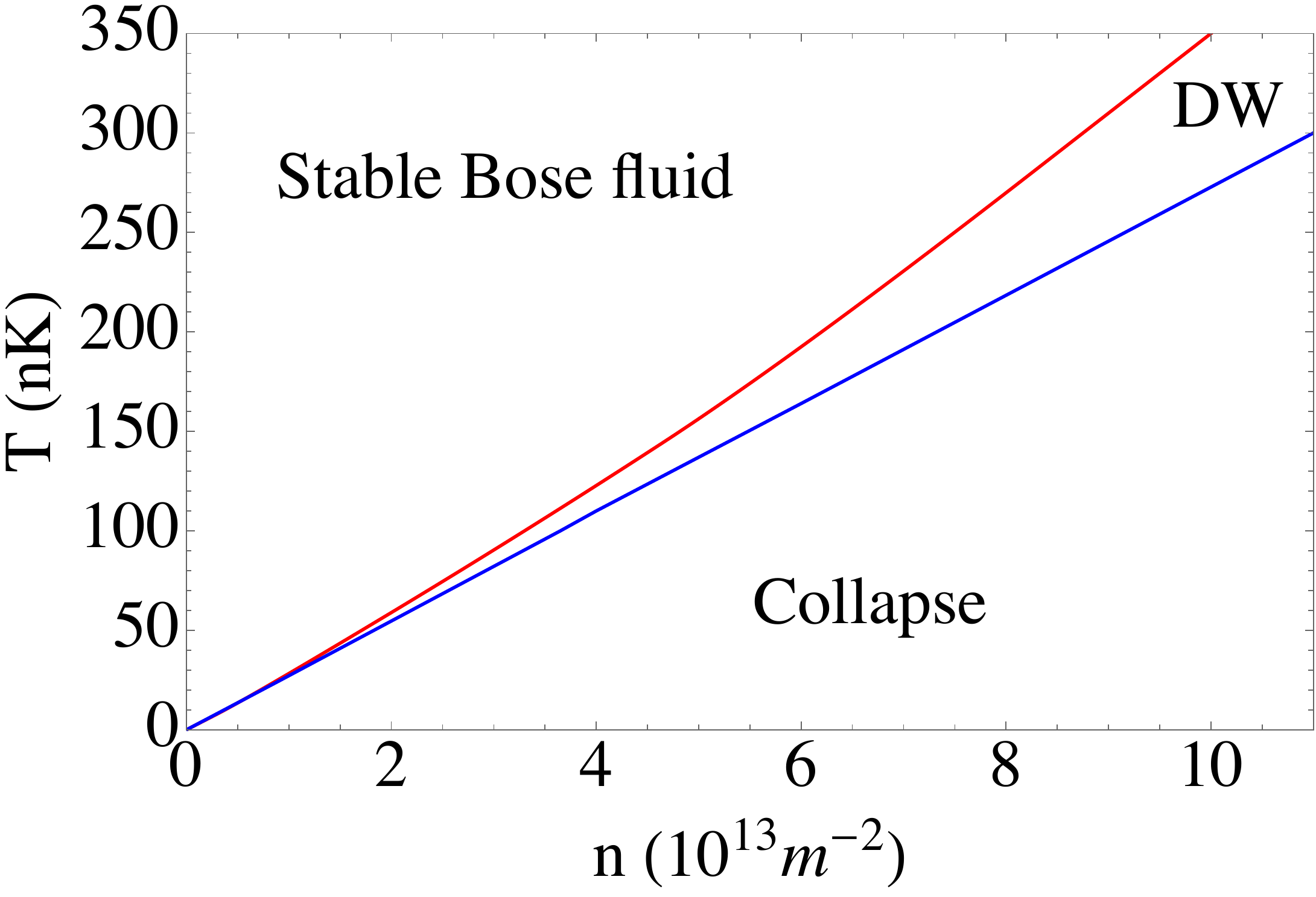}
	\end{minipage}
	\caption{(color online) Calculated T-n phase diagram for dipolar Bose gas at $\theta$ = 0 (left column) and $\theta=1$ (right column) $^{41}Rb^{87}Cs$ with dipole moment d=0.355D (top) and d=1.22D (middle); $^{166}Er$ (bottom) with $d=7\mu_b$. Density wave instability (DWI), red curve, occurs at low temperature and high density for $\theta$ = 0;. For $\theta$ =1, the system goes from stable Bose fluid phase to density wave instability and then to collapse (blue line) as temperature decreases and density increases.}
	\label{T-n}	
\end{figure}

Next, we construct a temperature-density (T-n) phase diagram for fixed dipole strengths $a_{dd}$, and tilt angles $\theta$.  For $\theta<\cos^{-1}{\frac1{\sqrt3}}$, as temperature decreases and density increases, the system goes through a transition from stable Bose fluid phase to a density wave (DW) phase. At $T=0$, there is a critical density below which there is no density wave instability, and the system is a BEC. For 
$\theta>\cos^{-1}{\frac1{\sqrt3}}$, as temperature decreases and density increases, the system goes through a transition from stable Bose fluid to a density wave phase to a collapse phase.  In Fig. \ref{T-n}, we show the calculated T-n phase diagrams, for $\theta=0$ and $\theta=1$, using parameters relevant to several physically realized dipolar boson systems, namely polar molecule $^{87}Rb^{133}Cs$ with electric dipole moment of 0.355 Debye and $a_{dd}=4586a_0$ (accessible in experiment) \cite{molony2014creation};  $^{87}Rb^{133}Cs $ with electric dipole moment of 1.22 Debye and $a_{dd}=93084a_0$  (the maximum value possible in experiments); $^{168}Er$ with magnetic dipole moment of $7\mu_b$ and  $a_{dd}=196a_0$ \cite{ferlaino-roton2018}. The sets of plots show that for larger dipolar interaction strength $a_{dd}$, the instability occurs at a larger density and lower temperature; compare for example, the plots for $^{87}Rb^{133}Cs $ with electric dipole moment of 1.22 Debye ($a_{dd}=93084a_0$) with $^{168}Er$ with magnetic dipole moment of $7\mu_b$ ($a_{dd}=196a_0$).
Density wave instability occurs at low temperature and high density for small tilt angle. For large tilt angle, the system goes from stable Bose fluid phase to density wave instability and then to collapse as temperature decreases and density increases; no discernible region of BEC appears at T=0 (as seen for $\theta =1$).

We have considered the effect of an additional short-range interaction $g$, originating from Van Der Waals interaction between atoms or molecules; this results in total interaction 
$V(q)=g+V_{dipole}(q)$. Within RPA, the main modifications are: a repulsive $g$ increases critical tilt angle, $\theta_c$ to a value larger than 0.955, while an {\it attractive} $g$ decreases it. 
This is because $\theta_c$ is now determined by the net short range interaction, that has contribution from the contact interaction, in addition to that contained in the dipole interaction. 
And adding a repulsive $g$ increases the critical $a_{dd}$ for instability, while an attractive $g$ decreases this. (See Supplemental Materials, Fig. S3~\cite{supp}).
 
\begin{figure}[h]
	\centering
	\includegraphics[width=0.9\linewidth]{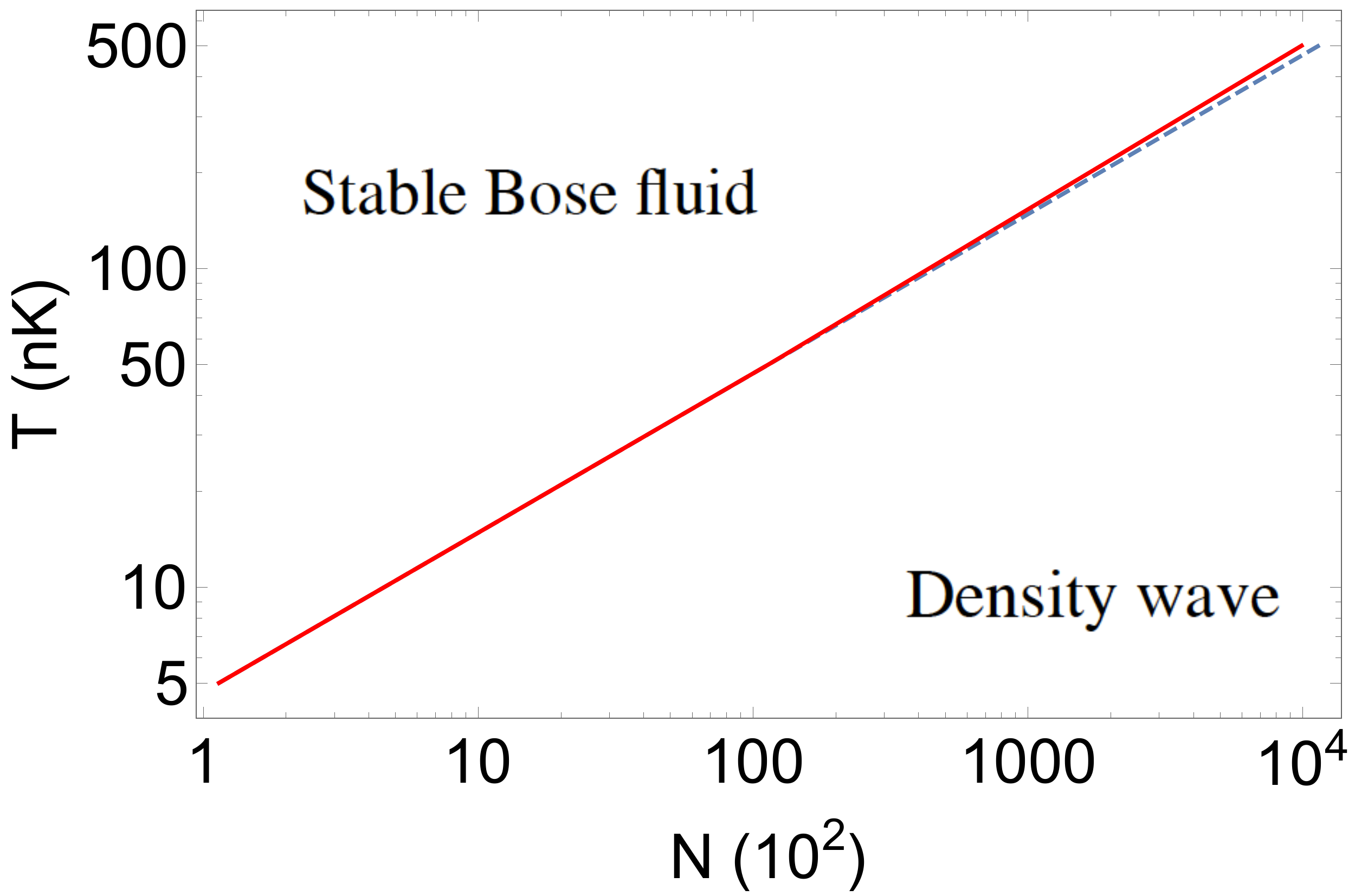}
	\caption{(color online) Calculated $T$-$N$ phase diagram for a system of trapped $^{166}Er$ at zero tilt angle. Here, $d=7\mu_B$ and trap parameter $\hbar w_r/k_B=0.6K$. The dashed line is for an ideal gas BEC phase transition, the solid line is the density wave instability line.}
	\label{trap}
\end{figure}

We briefly discuss possible effects of a trap given by a cylindrically symmetric harmonic potential $U(z,r)=1/2(w_zz^2+w_r r^2)$, where $z$ and $r$ are the 
axial and radial directions respectively. We explore instability in a quasi-2D trapped system with strong harmonic confinement in the $z$-direction, i.e.  $w_z\gg w_r$, by calculating the RPA response function 
$\chi$ within the local-density approximation (LDA).  We expect the instability to occur when the effective chemical potential, $\mu_{\text{eff}}(r)=\mu-\omega r^2/2$, is same as that of the uniform system at a certain temperature~\cite{mueller2000}; thus the local bare response function is the same as that of the uniform system.
We construct a T-N phase diagram, Fig. \ref{trap}, using the parameters for $^{166}Er$; N being the total number of particles. The density wave instability temperature $T_c$ is sightly above the ideal gas BEC temperature $T^0_{\text{BEC}}$. At $T_c$, the 2D trapped gas has density wave instability at $r=0$, where $\mu_{\text{eff}}$ is the largest, and begins to form stripe pattern starting from center of the trap. (See details in Supplemental Materials, Effect of a harmonic trap~\cite{supp}.) This behavior can be understood by noting that for trapped particles, condensation results in a huge increase in the central density of the cloud (a standard diagnostic of BEC)~\cite{mueller2000}.

We have shown that a broad understanding of the nature of phases and instabilities in a 2D dipolar Bose gas at finite temperature may be obtained on concurrently exploring 
tunable system parameters, namely, density, temperature, interaction strength and tilt angle. The presented phase diagrams provide different perspectives on the nature of the instabilities. 
We have used a finite-temperature version of RPA, and note that RPA is a well-established many-body method that has proved to be useful in describing collective modes and instabilities in quantum fluids.
At T=0, our finite-temperature RPA reproduces the BdG results, as expected~\cite{nozieres2018theory}.
Our approach and results may be of relevance generally to systems with long-range and anisotropic interactions, and thus of broader appeal.
The results may be compared with previous work in 3D with attractive contact interaction~\cite{mueller2000}, or dipolar interaction~\cite{bisset2011}, wherein possible long wavelength (q$\to$0) instabilities were studied. We note that density wave instability usually triggers a long-range order of the stripe phase. At finite temperature, enhanced fluctuation in 2D will destroy the long-range order, but a quasi-long-range order may survive. A phase transition belonging to the usual Berezinskii-Kosterlitz-Thouless universality class is expected; this has been studied in Monte Carlo simulation~\cite{filinov2010berezinskii,Bombin2019}. Thus, the $T_c$ curves discussed here, are in a strict sense, RPA instability lines. Accordingly, our $T$-$n$ phase diagram may need to be modified at low temperature; this is beyond the scope of RPA. Nevertheless, the RPA method does provide a picture of the instabilities of 2D dipolar boson system at finite temperatures that is expected to be qualitatively correct. 

\section{Acknowledgements}

We thank J. Boronat and G. Baym for useful discussions. We acknowledge funds from Institute for Complex Adaptive Matter. K. Q. acknowledges the hospitality of Aspen Center for Physics, where part of the work was done.

\bibliography{bibtex}

\end{document}


\pagestyle{empty}

\title{Supplemental materials for Finite Temperature Instabilities of 2D Dipolar Bose Gas at Arbitrary Tilt Angle}
\maketitle
\begin{center}
Pengtao Shen and Khandker Quader,\\
Department of Physics, Kent State University, Kent, OH 44242
\end{center}

\section{RPA diagrammatic expansion}
Diagrammatic expansion of the density-density correlation function, $\chi(q)$  in terms of the bare response function  $\chi_0$ is given below:

\begin{figure}[h]
	\centering		
	\includegraphics[width=\linewidth]{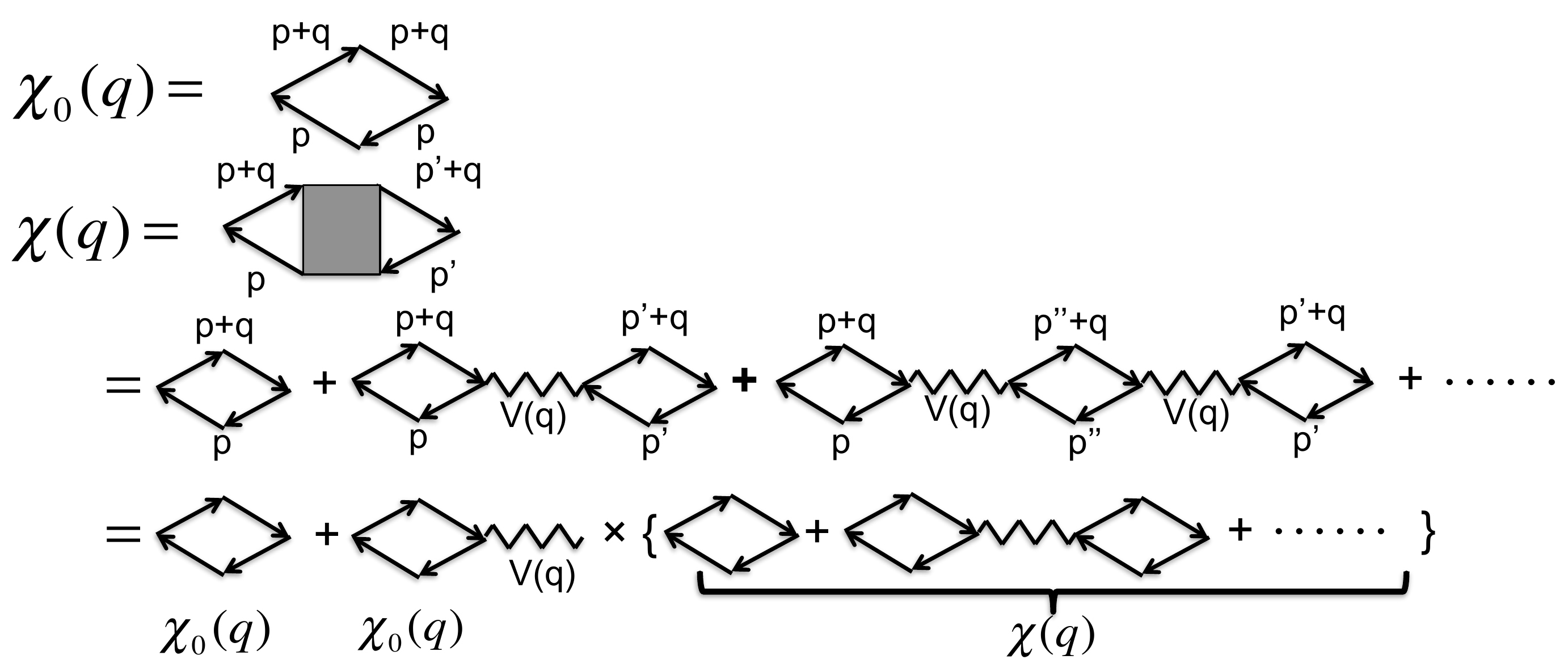}
	\caption{Feynman diagram expansion of density-density correlation function. $\chi_0$  is the bare response; $\chi(q)$ the total response function.  Wiggly lines denote interaction; 
	straight lines represent particles or holes.}
\end{figure}

\section{Finite temperature 2D boson bare bubble}

We derive the 2D boson bare bubble at finite temperature with chemical potential $\mu$. We start from the expression:

\begin{eqnarray}\label{barebubble}
\chi_0(q,\omega)=\int \frac{d \bk}{(2\pi)^d} \frac{f(k-q/2)-f(k+q/2)}{\hbar\omega-(\varepsilon_{k+q/2}-\varepsilon_{k-q/2})}
\end{eqnarray}
where $\varepsilon_k=\hbar^2 k^2/{2m}$ is the free particle kinetic energy, and $f(q)$ is the Bose distribution function, with chemical potential $\mu$.
We first shift $f(k\pm q/2)$ to $f(k)$ in Eq. (\ref{barebubble}) and then integrate over angles to obtain,
\begin{eqnarray}
\chi_0(q)=-\frac{2m}{\pi\hbar^2q}\int^{\frac q 2}_0 dk k \frac{1}{e^{\beta\varepsilon_k}e^{-\beta\mu}-1} \frac{1}{\sqrt{q^2-4k^2}}
\end{eqnarray}

In the $q\lambda_T\ll1$ limit,  $e^{\beta\varepsilon_k}=e^{\frac{{(k\lambda_T)}^2}{4\pi}}\simeq1+O(k\lambda_T)^2$. Integrating out k, we obtain
\begin{eqnarray}
\nonumber
\chi_0(q)&\simeq&-\frac{2m}{\pi\hbar^2q}\int^{\frac q 2}_0 dk k \frac{1}{e^{-\beta\mu}-1} \frac{1}{\sqrt{q^2-4k^2}}\\
&=&-\frac{m}{2\pi\hbar^2}\frac{1}{e^{-\beta\mu}-1}(1+O(q\lambda_T)^2)
\end{eqnarray}

In the $q\lambda_T\gg1$ limit, we obtain
\begin{eqnarray}
\nonumber
\chi_0(q)&\simeq&-\frac{2m}{\pi\hbar^2q^2}\int_0^{\infty} dk k \frac{1}{e^{\beta\varepsilon_k}e^{-\beta\mu}-1}\\
&=&-\frac{4nm}{\hbar^2q^2}
\end{eqnarray}\\

The plot below shows the bare response function for various temperatures.

\begin{figure}[h]
	\centering		
	\includegraphics[width=4.5in,height=3.25in]{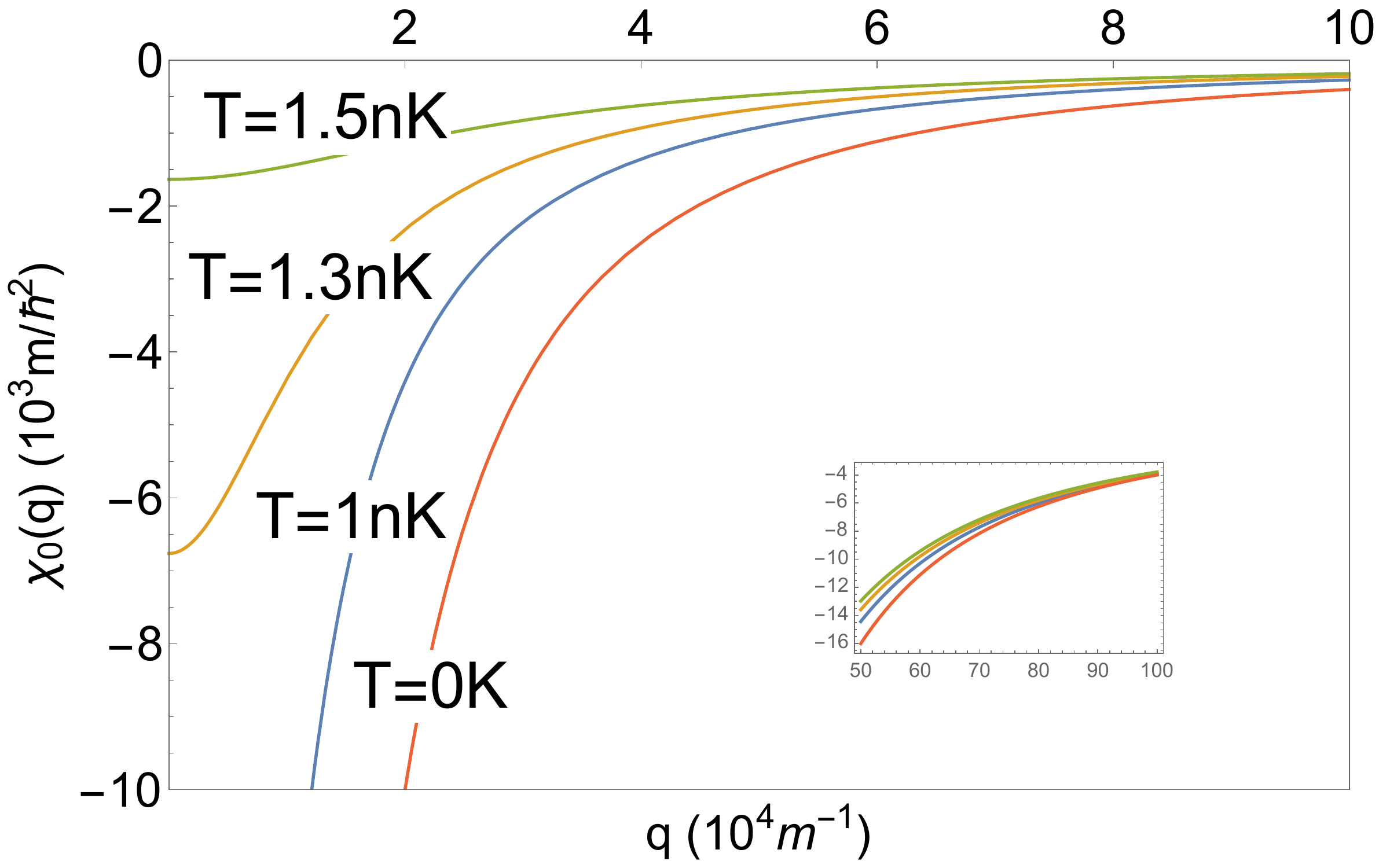}
	\caption{(color online) Bare response function calculated using mass of polar molecule $^{87}Rb^{133}Cs$ at density $n=10^{12}m^{-2}$ at varying temperatures. The inset is the response function at large momentum, showing that the curves asymptotically approach each other. }
\end{figure}

\newpage

\section{Additional Contact Interaction}

Plots of results for the cases of dipolar plus an additional repulsive and attractive contact interactions are shown below.

\begin{figure}[h]
	\centering
	\includegraphics[width=.36\linewidth]{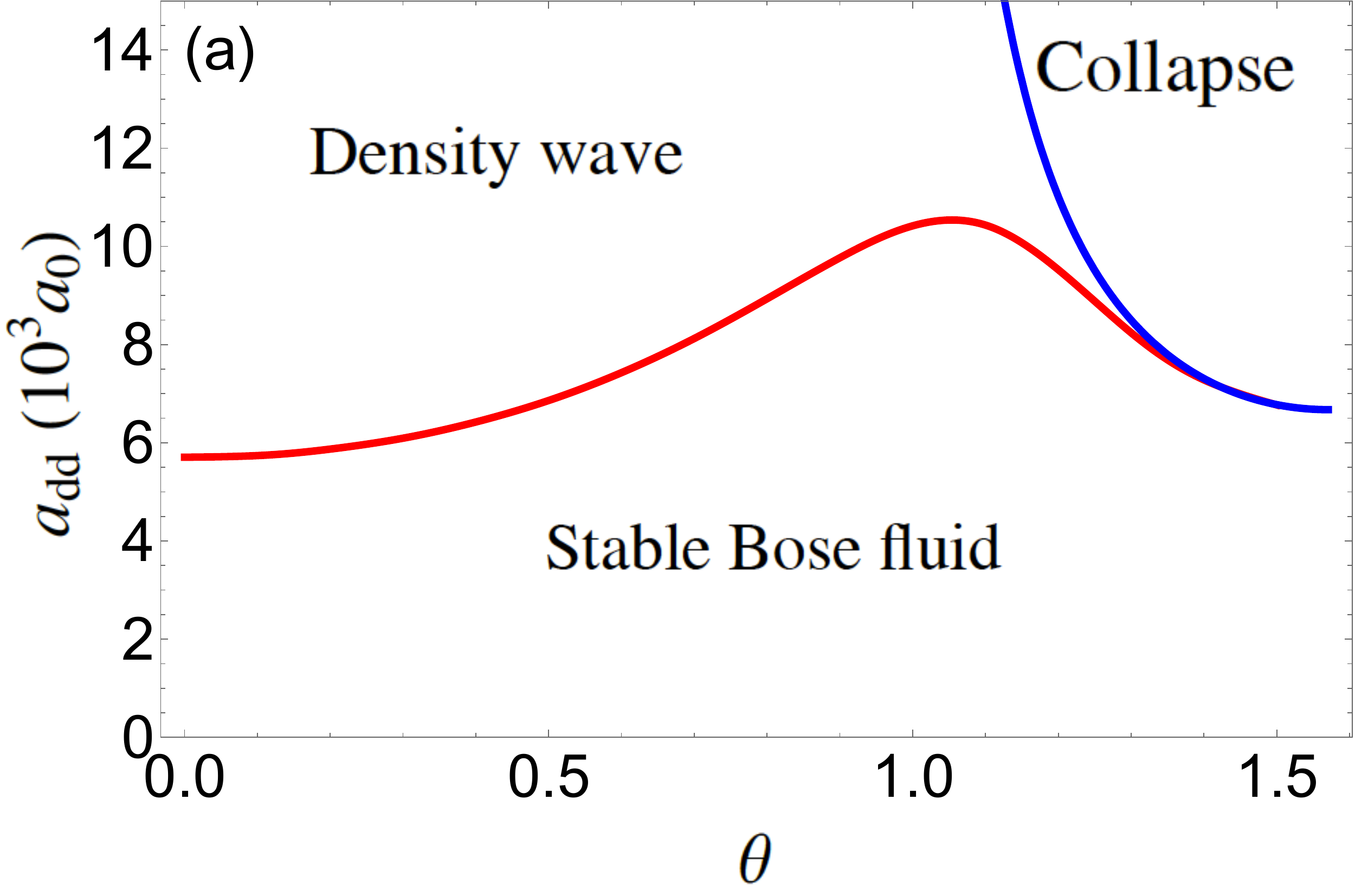}
	\includegraphics[width=.36\linewidth]{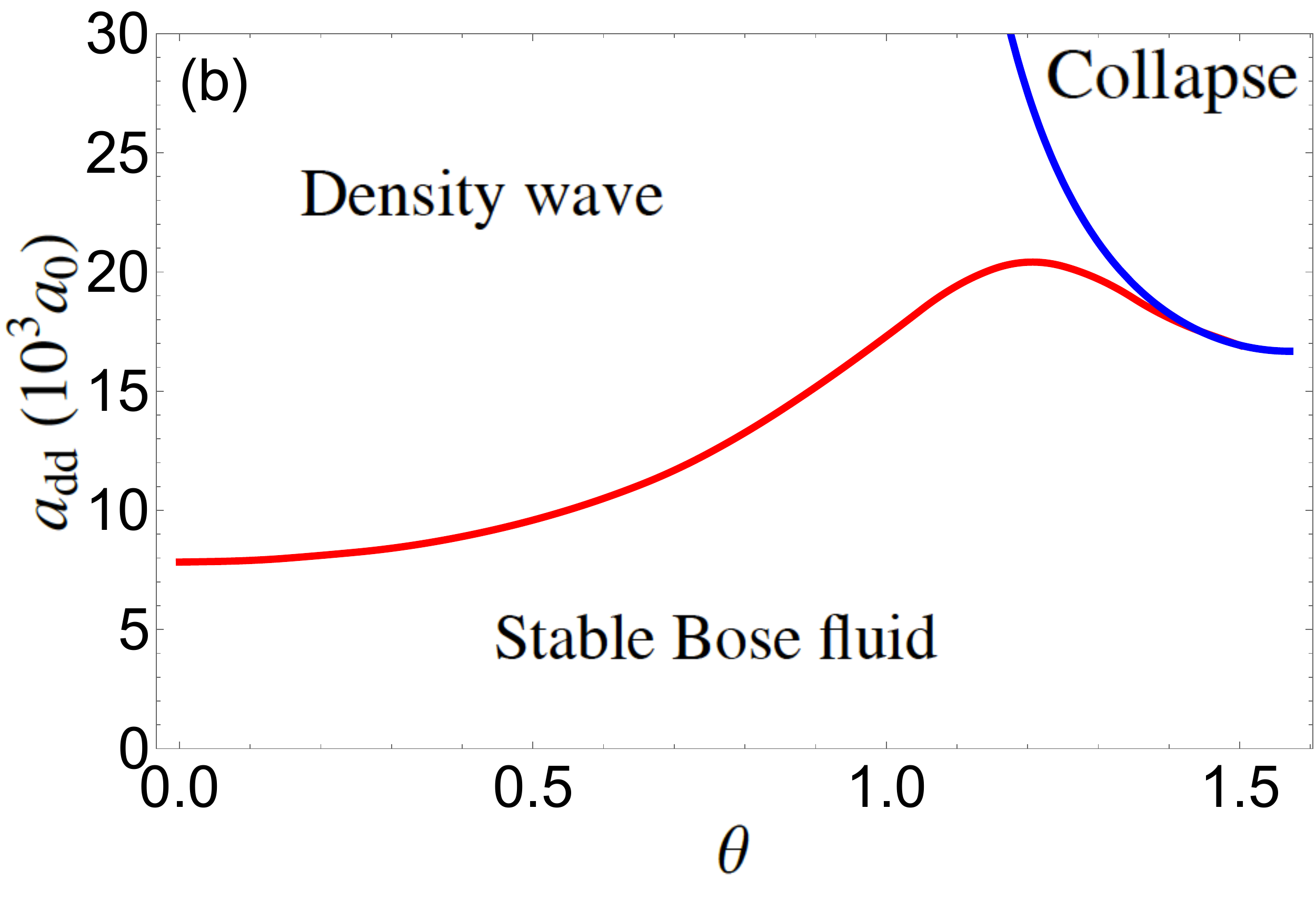}
	\includegraphics[width=.36\linewidth]{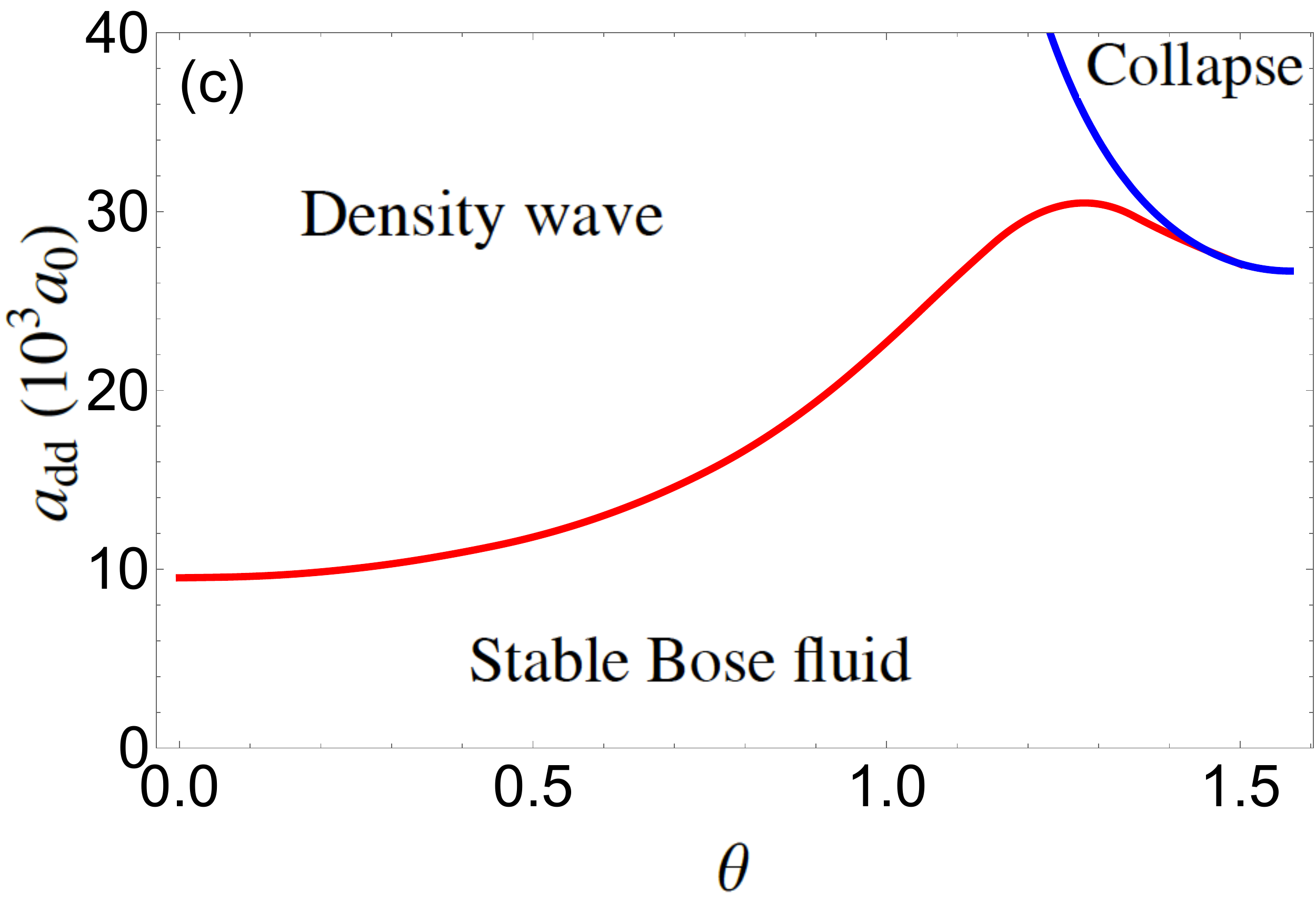}
	\includegraphics[width=.36\linewidth]{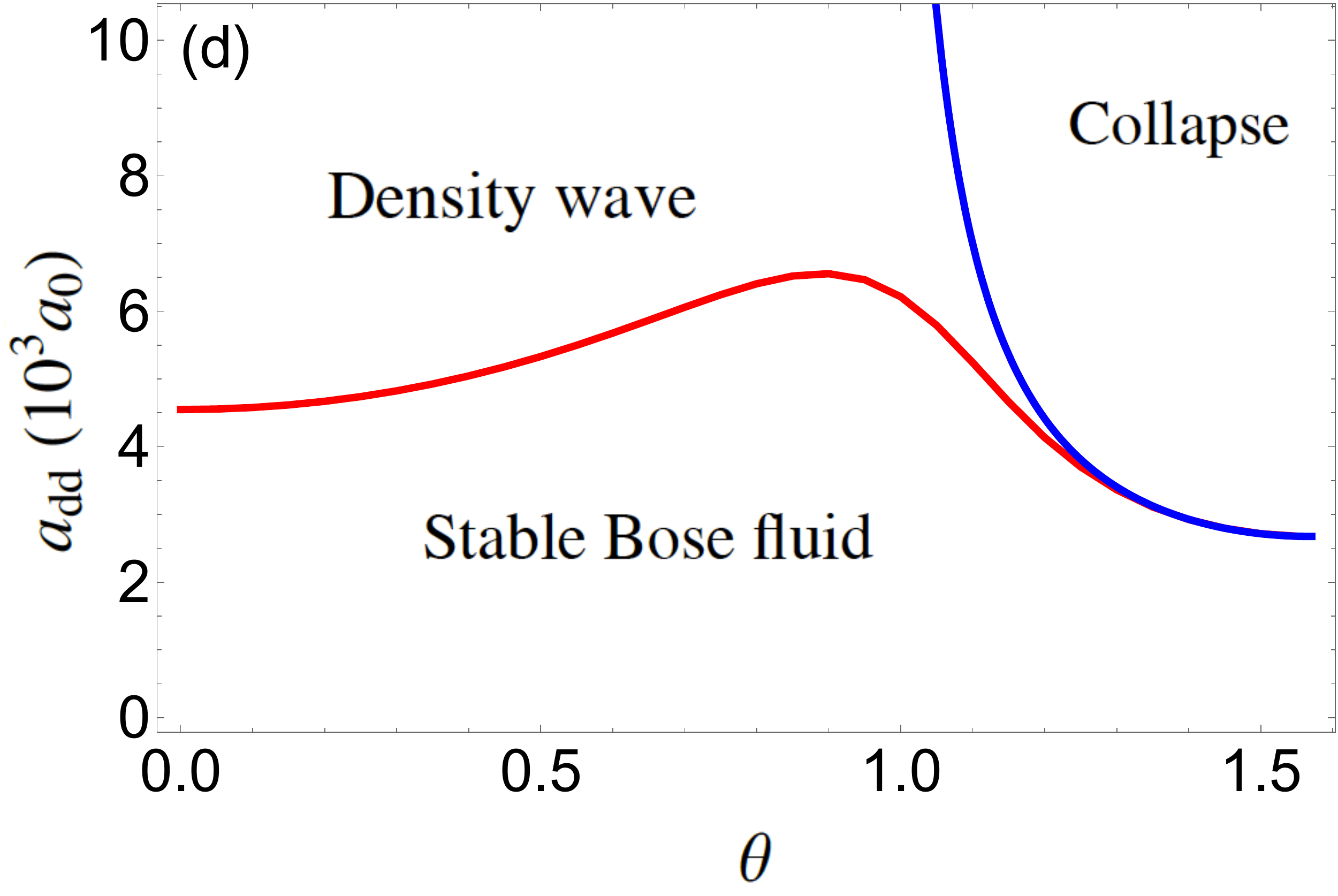}
	\caption{(color online) Plot showing the effect of adding an additional contact interaction, $g$. Calculated  $a_{dd}$ versus $\theta$ phase diagrams of polar molecule $^{87}Rb^{133}Cs$ at density $n=10^{12}m^{-2}$ at T=10 nK. $g=2\pi\frac{\hbar^2}{2m}a$ with  $a=0,0.5,1,-0.2$ for (a), (b), (c) and (d) respectively.}
	\label{g+ddi}
\end{figure}

\section{Effect of a harmonic trap}

A simple trap is in the form of cylindrically symmetric harmonic potential $U(z,r)=1/2(w_zz^2+w_r r^2)$, where z is the axial direction and r the radial direction. Changing the aspect ratio of the trap potential 
$\lambda=w_z/w_r$ gives different forms of the trap. The trap is pancake-shaped when $\lambda>1$  and cigar-shaped when  $\lambda<1$. Due to the anisotropic and long range feature of DDI, the trap form plays an important role in the stabilization of dipolar gas system~\cite{nature-dg-stable,bisset2011}. Here we consider a quasi-2D trapped system with strong harmonic confinement in the z-direction,  
e.g., $w_z\gg w_r$.  

As in case of the uniform 2D system, we calculate the response function $\chi$ to determine the instability. $\chi$ is calculated within the local-density approximation (LDA), treating the gas locally as at uniform density with chemical potential potential $\mu_{\text{eff}}(r)=\mu-1/2 w_r r^2$.  For a 2D trapped ideal boson gas with $k_BT\gg\hbar w$, the density profile is well described by the local density approximation,
\begin{eqnarray}
n(r)=\int \frac{d^2p}{(2\pi)^3}\frac 1{e^{\beta[\varepsilon_p+1/2w_r r^2-\mu]-1}}+n_0e^{-r^2/d^2_r}
\end{eqnarray}
where $\varepsilon_p=p^2/2m$, $\mu$ is the chemical potential, $n_0$ is condensate density at $r=0$. Below the BEC critical temperature $T_c$, condensation exist and $n_0\ne0$, $\mu$=0, Above  $T_c$, there is no condensation, and $n_0=0$, $\mu<0$.
We can obtain the total number of particles by carrying out a spatial integral $\int dr$ of Eq.(5),
\begin{eqnarray}
N=\begin{cases}
(\frac{1}{\beta\hbar w})^2g_2(e^{\beta\mu})  &T<T_c\\
(\frac{1}{\beta\hbar w})^2\zeta(2)+(\frac{\pi}{\hbar w})^{\frac12}n_0     &T>T_c
\end{cases}
\end{eqnarray}
where $g_n(x)$ is Polylogarithm function and $\zeta(x)$ is Riemann zeta function, $g_n(1)=\zeta(n)$. By taking $\mu=0$, $n_0=0$ and $\zeta(2)=\pi^2/6$, we find BEC temperature for ideal Boson gas in trapped 2D geometry is $T^0_{\text{BEC}}=\sqrt{6N}\hbar w/k_B\pi$. At $T^0_{\text{BEC}}$, the density of excited particle follows $n(r)=-\frac1{\lambda^2_T}\log(1-e^{-\beta w_r r^2/2})$. The density at the center (r=0) is infinite, however the total number of particles, $N$ remain finite upon performing the integral. With the presence of (repulsive) interaction, mean-field solutions suggest critical BEC point may not exist~\cite{hadzibabic2008trapped}. One might expect a BKT type of transition near the ideal gas BEC temperature $T^0_{\text{BEC}}$~\cite{holzmann2007superfluid}.

\bibliography{bibtex}